\documentclass[%
 aip,
 amsmath,amssymb,
 reprint,%
]{revtex4-1}

\usepackage{amsmath}
\usepackage{amssymb}
\usepackage{amsfonts}
\usepackage{array}
\usepackage{graphicx}
\usepackage{mathrsfs}
\usepackage{multirow}
\usepackage{siunitx}
\usepackage{color}
\usepackage{subfigure}
\usepackage{comment}

\newcommand{\feff}{\ensuremath{F_\text{eff}}}
\newcommand{\veff}{\ensuremath{V_\text{eff}}}
\newcommand{\kb}{\ensuremath{k_\text{B}}}
\newcommand{\phihs}{\ensuremath{\phi_\text{HS}}}
\newcommand{\bea}{\begin{eqnarray*}}
\newcommand{\eea}{\end{eqnarray*}}
\newcommand{\beao}{\begin{eqnarray}}
\newcommand{\eeao}{\end{eqnarray}}

\begin{document}

\title{Effective potentials induced by mixtures of patchy and hard co-solutes}
\author{Philip H. Handle}
\affiliation{%
Institute of Physical Chemistry, University of Innsbruck, Innrain 52c,
A-6020 Innsbruck, Austria
}
 \affiliation{ 
Department of Physics, "Sapienza" University of Rome, Piazzale A. Moro 5, I-00185 Rome, Italy
}

\author{Emanuela Zaccarelli}%
\affiliation{ 
Department of Physics, "Sapienza" University of Rome, Piazzale A. Moro 5, I-00185 Rome, Italy
}
\affiliation{%
CNR Institute for Complex Systems, Uos "Sapienza", Piazzale A. Moro 2, I-00185 Rome, Italy
}
\author{Nicoletta Gnan}
 \email{nicoletta.gnan@roma1.infn.it}
\affiliation{ 
Department of Physics, "Sapienza" University of Rome, Piazzale A. Moro 5, I-00185 Rome, Italy
}%
\affiliation{%
CNR Institute for Complex Systems, Uos "Sapienza", Piazzale A. Moro 2, I-00185 Rome, Italy
}%

\date{\today}

\begin{abstract}
The addition of co-solutes to colloidal suspensions is often employed to induce tunable depletion interactions. 
In this work we investigate effective colloidal interactions arising from binary co-solute mixtures of hard spheres and patchy particles.
By changing the relative concentration of the two species, we show that the resulting effective potential \veff \  continuously changes from the one obtained for a single-component hard sphere co-solute to that mediated by the single-component patchy particle co-solute.
Interestingly, we find that, independently of the relative concentration of the two components,  the resulting \veff \  is additive, i.e., it  is well-described by the linear combination of the effective interactions mediated by respective pure co-solutes. 
However, a breakdown of the additivity occurs when the co-solute mixture is close to the onset of a demixing transition.
These results represent a step forward in understanding and predicting colloidal behaviour in complex and crowded environments and for exploiting this knowledge to design targeted colloidal superstructures.
\end{abstract}

\maketitle

\section{Introduction}
The archetypal example of depletion interactions emerge when hard-sphere colloids are dispersed in a solvent with non-adsorbing polymers (the so-called co-solute). In such systems the presence of the polymers give rise to an entropy-driven effective attraction between the colloids. More than 60 years ago Asakura and Oosawa studied depletion interactions when an idealized polymer is considered as a co-solute.~\cite{asakura1958interaction}
In this case the depletion interaction can be derived analytically~\cite{asakura1958interaction,vrij1976polymers}
and the Asakura-Oosawa model has since become a widely used reference system.~\cite{binder2014perspective}
Analytic treatment of colloidal suspensions is however limited to a few specific cases and the entropic depletion interaction is part of a larger family of effective interactions mediated by different kinds of co-solutes.
Such effective interactions are ubiquitous in nature~\cite{likos2001effective,marenduzzo2006depletion,lekkerkerker2011colloids} and can induce phase separation, enhance crystallization and gelation, or give rise to different types of arrested states.
The significant impact on the behavior of colloidal systems makes the deliberate control of effective interactions highly desirable as this would allow for the design of specific materials, which can be exploited in fields spanning catalysis, photonics, sensor and filter technology, energy storage, medicine, and food science.~\cite{velev1999in,shipway2000,man2005experimental,wang2010targeting,armstrong2015artificial,lindquist2017interactions,assenza2019soft}

Colloidal effective interactions can be controlled by changing the relative size and the density of the co-solute, but also by choosing the proper co-solute--colloid and co-solute--co-solute interactions.
For instance, the model studied by Asakura and Oosawa can be modified by using different types of colloids
(e.g., hard spherocylinders,~\cite{lekkerkerker1994phase,bolhuis1997numerical}
or soft spheres~\cite{lonetti2011ultrasoft,rovigatti2015soft}),
different types of co-solutes
(e.g.,
hard spherocylinders,~\cite{roth2002depletion} star-polymers,~\cite{mayer2009multiple,mahynski2013phase}
or self-propelling particles~\cite{das2014phase}),
or different types of both
(e.g.,
hard spherocylinders with a hard sphere co-solute,~\cite{li2005depletion}
hard rods with a binary Lennard-Jones co-solute~\cite{sapir2014origin},
or microgel mixtures~\cite{bergman2018new}).
Also polydisperse systems have been investigated~\cite{kalyuzhnyi2003phase}
as well as systems in which the co-solute particles are close to a second-order critical point thus inducing tunable long range interactions among the colloids.~\cite{Casimir2012JCP,CasimirGnan2014,maciolek2018collective}
The case of specific interest for the present work is the one in which the co-solute self-assembles into a specific structure.

In studying patchy particles~\cite{pawar2010fabrication,bianchi2011patchy,bianchi2017limiting} as co-solutes some of us have recently shown that, if the co-solute reversibly self-assembles into non-adsorbing polymer chains, attractive oscillating two-body interactions between colloids arise.~\cite{garcia2017effective}
This is an especially intriguing result in light of several other studies, which have shown that the self-assembly of one-component systems can be controlled through designed oscillating interactions,~\cite{torquato2009inverse, cohn2009algorithmic,jadrich2015equilibrium,lindquist2017interactions}
leading, for example, to the formation of specific crystals~\cite{rechtsman2006self, edlund2011designing, marcotte2013communication, jain2014dimensionality,engel2015computational} and
quasicrystals.~\cite{ryltsev2015self,engel2015computational,metere2016smectic,damasceno2017non}
Yet, it is unknown how to design experimental systems to give rise to such potentials.
Encouraged by the results emerging in colloidal dispersions with a self-assembling co-solute,~\cite{garcia2017effective} in this work we explore how the effective interaction between the colloids can be tuned by using co-solute binary mixtures, where one of the two components is capable of self-assembly.
Specifically, we investigate mixtures of hard-spheres and patchy particles of equal sizes for different concentrations of the two species.
In using these systems, we study whether the resulting effective potentials can be predicted from those obtained for the respective single-component co-solutes.

The paper is organized as follows:
In Sec.~\ref{sec:modelsmethods} we introduce the  model system and the applied simulation techniques.
In Sec.~\ref{sec:mono} we discuss the effective potentials mediated by the single-component co-solutes.
Sec.~\ref{sec:mix} provides an analysis of the structural properties of the co-solute mixtures in the absence of the two colloids and in Sec.~\ref{sec:lincomb} we discuss the resulting effective potentials mediated by the co-solute mixtures.
In Secs.~\ref{sec:mono}--\ref{sec:lincomb} the patchy co-solute particles considered have two attractive patches (2P).
In Sec.~\ref{sec:3P-HS} we present the same analysis, but instead of the 2P patchy co-solute, we use patchy co-solute particles with three attractive sites (3P).
Finally, in Sec.~\ref{sec:conclusions} we summarize all results and present our conclusions.

\section{Models and Methods}
\label{sec:modelsmethods}
\begin{figure}
  \includegraphics[width=1.0\linewidth]{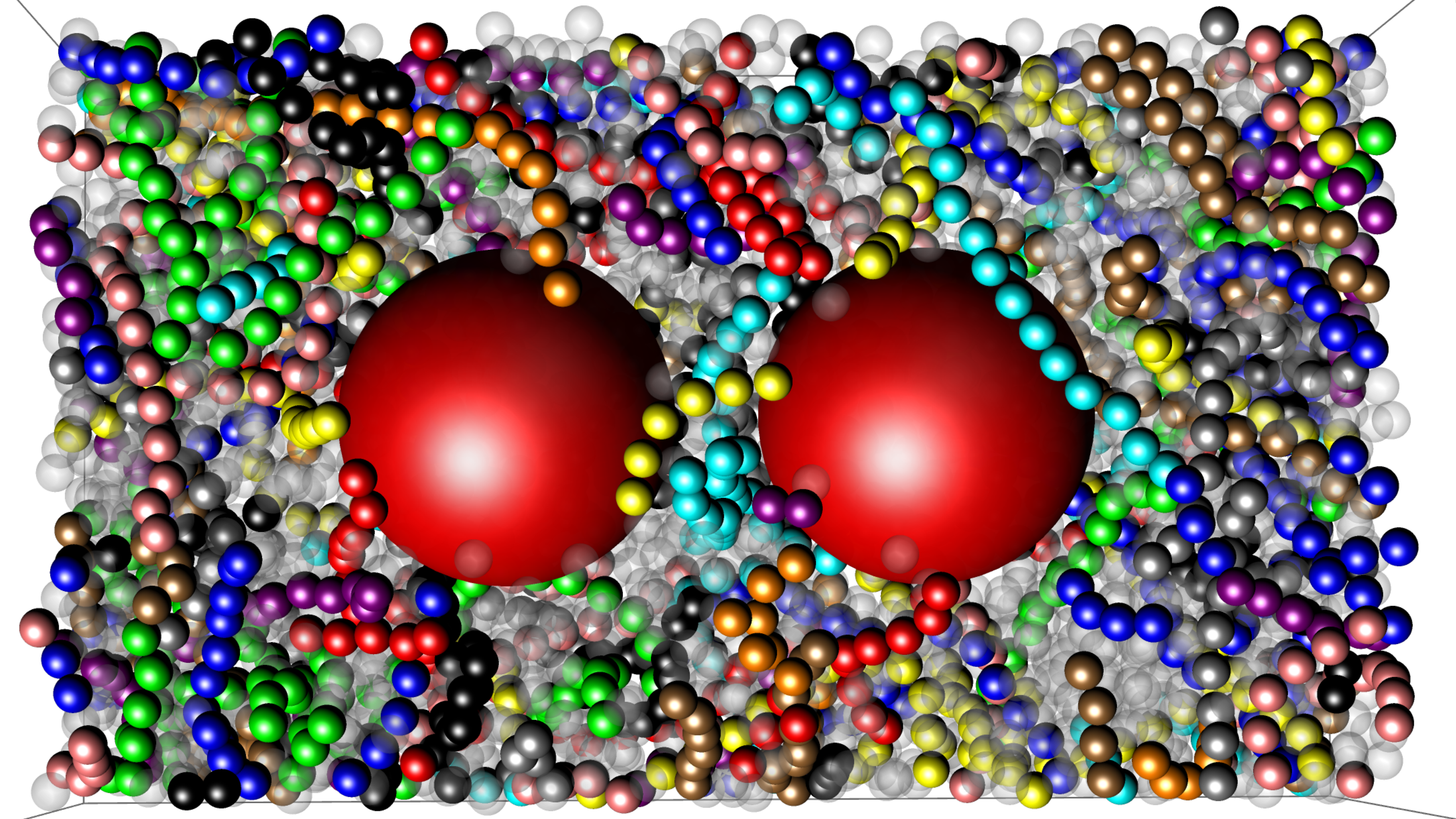}
    \caption{Snapshot of two HS colloids (red, $\sigma_c=10\sigma$) in a binary co-solute made of $50\%$ HS (transparent spheres, $\sigma=1$) and $50\%$ 2P particles ($\sigma=1$) at a total packing fraction $\phi=0.262$.
    At low temperature the 2P particles self-assemble into chains, highlighted in the snapshot with different colors.  }
\label{fig:snap}
\end{figure}
We perform $NVT$ Monte Carlo (MC) simulations of two colloids immersed in a binary mixture composed of hard-spheres (HS) and patchy particles of valence two (2P), or three (3P) at a packing fraction $\phi=0.262$ to evaluate the effective interactions at different species concentrations.
 A representative snapshot for the HS-2P co-solute mixture is shown in Fig.~\ref{fig:snap}.
In all cases, colloid--co-solute interaction is provided only by excluded volume and is of hard-sphere type. The size of all different co-solute particles (HS, 2P, 3P) is the same and  it is set to $\sigma=1$, while the size of the two colloids is $\sigma_c=10\sigma$.
Consequently, $\sigma$ is used as the unit of length throughout this article.

The patchy co-solute particles (2P or 3P) are modelled as hard spheres decorated with two or three attractive sites (patches), respectively.
Site-site interaction is treated using the Kern-Frenkel potential,~\cite{kern2003fluid,SimPatchyPart2018,Gnan2019Proteins} which originally was introduced by Bol:~\cite{bol1982monte}
\begin{eqnarray}
V_{\rm KF}(r_{ij}, \hat{n}^{\alpha}_i, \hat{n}^{\beta}_{j}) =V_{\rm SW}(r_{ij})\cdot f(\hat{n}^{\alpha}_i, \hat{n}^{\beta}_j).
\label{eq:KF}
\end{eqnarray}
This potential consists of a square-well attraction 
\begin{eqnarray}
V_{\rm SW}(r_{ij}) = 
\begin{cases}
\infty & {\rm if} \quad r_{ij} \leq \sigma \\[0.5em]
-\varepsilon & {\rm if} \quad \sigma < r_{ij} \leq \sigma+\delta \\[0.5em]
0 & {\rm if} \quad r _{ij}> \sigma+\delta
\end{cases}
\label{eq:SW}
\end{eqnarray}
of depth $\varepsilon$ (the unit of energy) and width $\delta$ modulated by an angular function, which depends on the mutual orientation of the patchy particles.
Specifically, two patches $\alpha$ and $\beta$ on particles $i$ and $j$ only attract each other when the center-to-center vector $\vec{r}_{ij}$ lies inside the cones of the two patches, i.e.,
\begin{eqnarray}
f(\hat{n}^{\alpha}_i, \hat{n}^{\beta}_j)= 
\begin{cases}
1  & {\rm if} \quad 
 \begin{cases}
 \quad \hat{r}_{ij}\cdot \hat{n}^{\alpha}_{i} > \cos(\theta_{\textrm{max}}) \\[0.5em]
 \quad \hat{r}_{ji}\cdot \hat{n}^{\beta}_{j} > \cos(\theta_{\textrm{max}}) \\[0.5em]
\end{cases}\\
0 & {\rm otherwise},
\end{cases}
\label{eq:angularf}
\end{eqnarray}
where the unit vector identifying, for instance, the orientation of patch $\alpha$ on particle $i$ is $\hat{n}^{\alpha}_{i}$.
The volume available for bonding is controlled by $\cos(\theta_{\textrm{max}})$. For all simulations we have fixed $\delta=0.119 \sigma$ and  $\cos(\theta_{\textrm{max}})=0.894717$ ensuring the single-bond per patch condition. 

After fixing the temperature $T$ and the packing fraction $\phi$, we evaluate the effective potentials \veff \ 
for different concentrations of the two co-solutes. For most simulations, we employ a parallelepipedal simulation box with lengths $L_x=50\sigma$ and $L_y=L_z=21\sigma$. In some cases we have simulated a smaller number of particles using a simulation box of size $35\sigma\times 20\sigma\times 20\sigma$. The box geometry always guarantees that the maximal surface-to-surface distance $r$ of the colloids is larger than the distance at which \veff{} goes to zero. Periodic boundary conditions are applied in all directions.
We employ a standard umbrella sampling technique~\cite{Casimir2012JCP,CasimirGnan2014, garcia2017effective} to evaluate the probability $P(r)$  of finding the two colloids at distance $r$ and we extract the effective interaction from the relation
\begin{equation}
    \beta \veff (r)=-\ln{\left(\frac{P(r)}{P(\infty)}\right)}=-\ln{g(r)},
\end{equation}
where $\beta=1/\kb T$, \kb{} is Boltzmann's constant and $g(r)$ is the two-body radial distribution function of the colloids.

Despite the efficiency of the method, which allows us to uniformly sample all distances between the colloids, we cannot directly separate the contributions of the two co-solute species to the total effective interaction. In order to obtain this information, we calculate the effective force \feff{} acting on the colloids in the following way:~\cite{wu1999monte,gnan2012properties}
at each MC simulation step, virtual displacements of size $\Delta r$ are performed by the two colloids. These displacements can be positive or negative. The probability that such a displacement leads to at least one collision is
\begin{equation}
    P_{\textrm{overlap}}(\Delta r)=\langle 1- e^{-\beta\Delta\phihs}\rangle.
\end{equation}
Here $\Delta\phihs$ is the change in the colloid--co-solute excluded volume interaction due to the virtual displacement $\Delta r$.
$\Delta\phihs=0$ when the virtual move results in no collision (the contribution to $P_{\textrm{overlap}}(\Delta r)$ is consequently 0) and $\Delta\phihs$ becomes infinite when the virtual move leads to at least one overlap (the contribution to $P_{\textrm{overlap}}(\Delta r)=1$ in this case).
This enables us to write \feff{} as:
\begin{align}\label{eq:virtual}
  \beta F^{i}_{\textrm{eff}}(r)= -\lim_{\Delta r \to 0^+}\frac{P^{i}_{\textrm{overlap}}(\Delta r)}{\Delta r}-\lim_{\Delta r \to 0^-}\frac{P^{i}_{\textrm{overlap}}(\Delta r)}{\Delta r}.
\end{align}
For a given colloid separation $r$ an imbalance of the overlap probability at positive $\Delta r$ ($\lim_{\Delta r \to 0^+}$) and the overlap probability at negative $\Delta r$ ($\lim_{\Delta r \to 0^-}$) gives rise to an effective force.
If the two probabilities are equal, \feff{} is zero. Notably, this way of calculating \feff{} allows us to calculate the total \feff{} as well as the individual contributions stemming from the HS and the patchy particles, respectively, by considering only overlaps of the colloids with one or the other co-solute species. In Eq.~\ref{eq:virtual} this is indicated by the index $i$, which can be tot, HS, or 2P (or 3P, respectively).

We note that the ratio $P^{i}_{\textrm{overlap}}(\Delta r)/ \Delta r$ 
is insensitive to the magnitude of the displacement as long as $\Delta r$ is small enough.
Previous works have shown that an optimal choice of magnitude provides an average collision probability of about $5\% - 15\%$.\cite{wu1999monte,gnan2012properties}
In this work we aimed for a collision probability of the whole mixture of $\approx10\%$, which implies that the collision probability for each species is lower than this value and depends on the relative amount of the species in the mixture.

Finally, we also perform MC simulations of the binary mixture of co-solutes, i.e., in the absence of the two colloids, to investigate its structural properties. To this end, we have focused on a smaller system ($N=1315$) in a cubic simulation box at the same $T$ and $\phi$ as used for the calculation of \veff{}.  From these simulations we extract the chain-length distribution $P(l)$, i.e., the probability $P$ of finding a chain of length $l$. In addition, we calculate the partial radial distribution function $g_{ij}(r)$ with respect to species $i$ and $j$ and the concentration-concentration structure factor which is an indicator of the demixing transition.
The latter is defined as~\cite{Thornton1970, ZaccaPRL2008,PriegoJCP139}
\begin{equation}
S_{cc}(q)=x^2_{2}S_{11}(q)-2x_1x_2S_{12}(q)+x^2_{1}S_{22}(q),
\end{equation}
where  $x_i$ is the concentration of the species $i$ and
\begin{equation}
S_{ij}(q)=\frac{1}{\sqrt{N_iN_j}}\sum_{i,j} e^{i\vec{q}\cdot(\vec{r_{i}}-\vec{r_{j}})}
\end{equation}
is the partial structure factor between species $i$ and $j$.

\section{Results:}

\subsection{Effective potentials induced by single-component co-solutes}\label{sec:mono}
\begin{figure*}[ht!]
\centering
\includegraphics[width=1.0\textwidth]{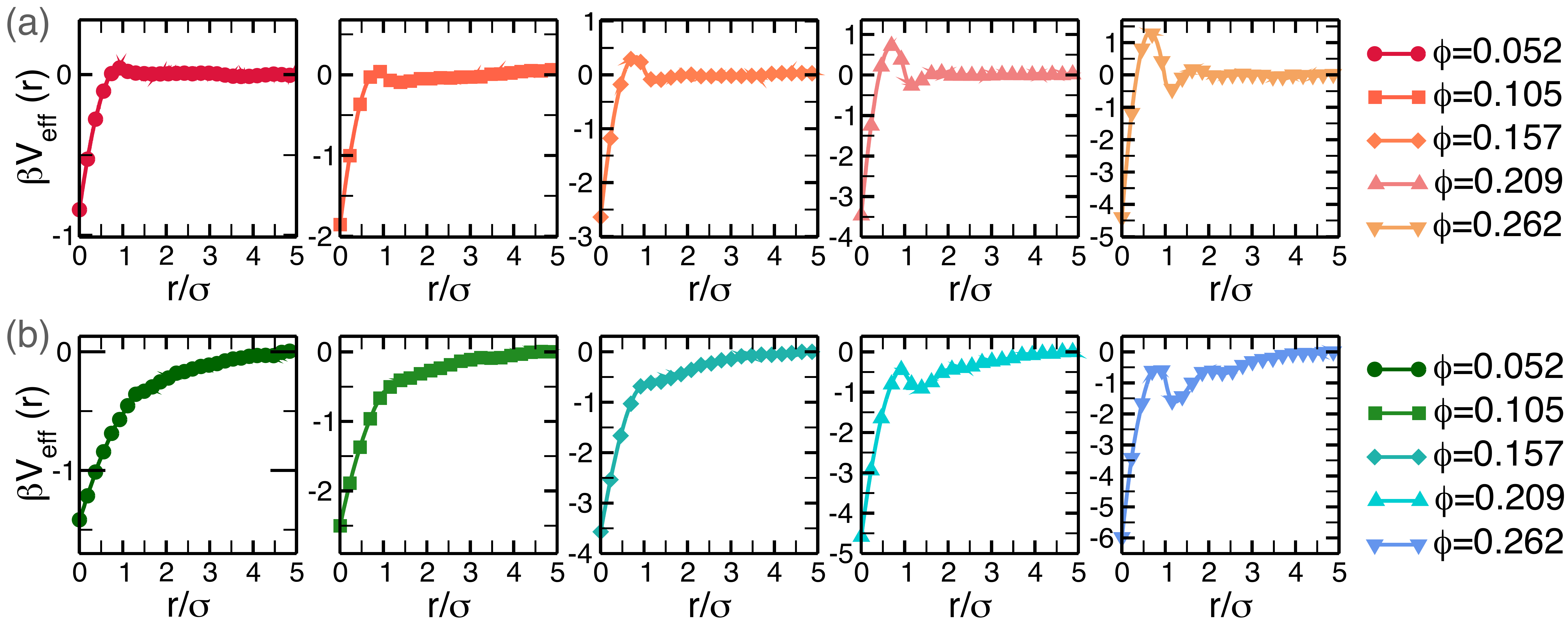}
\caption{(a) Effective potentials generated by a single-component co-solute of HS particles at different packing fractions $\phi$. (b) Effective potentials generated by a single-component co-solute of 2P particles at $T=0.1$ and different $\phi$.}
\label{fig:VeffMono}
\end{figure*}

We start our discussion by reporting the effective interactions arising between two large HS colloids immersed in a single-component co-solute (either HS or 2P particles). This preliminary study is fundamental to understand and possibly predict the effective interactions mediated by mixtures of these two components. Fig.~\ref{fig:VeffMono}(a) shows \veff{} mediated by a co-solute of HS type as a function of $\phi$. All potentials are characterized by an attractive well for $r/\sigma<1$, which results from the depletion of the co-solute between the colloids. In addition, on increasing $\phi$ the oscillatory character of $V_{\textrm{eff}}$ increases, signalling the emergence of correlations between the colloids due to the underlying structure of the co-solute.\cite{LekkerkerkerPhysA1995, DijkstraPRE59, DickmanJCP1997, RothPRE62}

A different scenario occurs for the co-solute of 2P particles, which is shown in Fig.~\ref{fig:VeffMono}(b). 
Here, we focus on a low temperature value, i.e., $T=0.1$, for which 2P particles are known to reversibly self-assemble into chains, whose lengths depend on the control parameters $T$ and $\phi$. In fact, the high-$T$ region is less interesting in terms of depletion interactions, since on increasing $T$ the bonding energy of the 2P particles becomes much smaller than the thermal energy, meaning that the particles behave essentially as HS.~\cite{garcia2017effective}

At low $\phi$ the effective potential between the large colloids is completely attractive and, differently from the standard depletion potentials, goes to zero at $r\approx 4\sigma$. The long-range nature of the resulting potential can be understood in terms of the chains formed by the co-solute, because the chains now act as a depletant with an effective diameter mediating the colloid-colloid interaction.
On increasing $\phi$, similar oscillations as found in the effective potentials mediated by the HS co-solute appear,  but the effective potentials for the 2P co-solute remains attractive at all distances. This has been explained in terms of the efficiency of 2P in ordering in the confined space between the colloids.~\cite{garcia2017effective} In particular, it has been shown that in between the two colloids 2P particles are correlated in orientation, reminiscent of nematization, in order to form bonds with other particles thus satisfying both energy and entropy constraints. The resulting minima and maxima of the oscillations  in \veff{} are roughly placed at distances of the order of the the size of the depletant as in the case of the effective potential generated by HS co-solute.
We stress that the attractive oscillations shown in Fig.~\ref{fig:VeffMono}(b) are a peculiarity of the reversibly self-assembling co-solute and have not been observed in the case of effective interactions mediated by irreversibly bound polymer  chains,~\cite{ZhangPolymerChains2016} since no competition between bonding and excluded volume takes place in that case.

\subsection{Characterization of the binary co-solute mixture without colloids}\label{sec:mix}

\begin{figure*}[tbh]
\centering
  \includegraphics[width=0.9\linewidth]{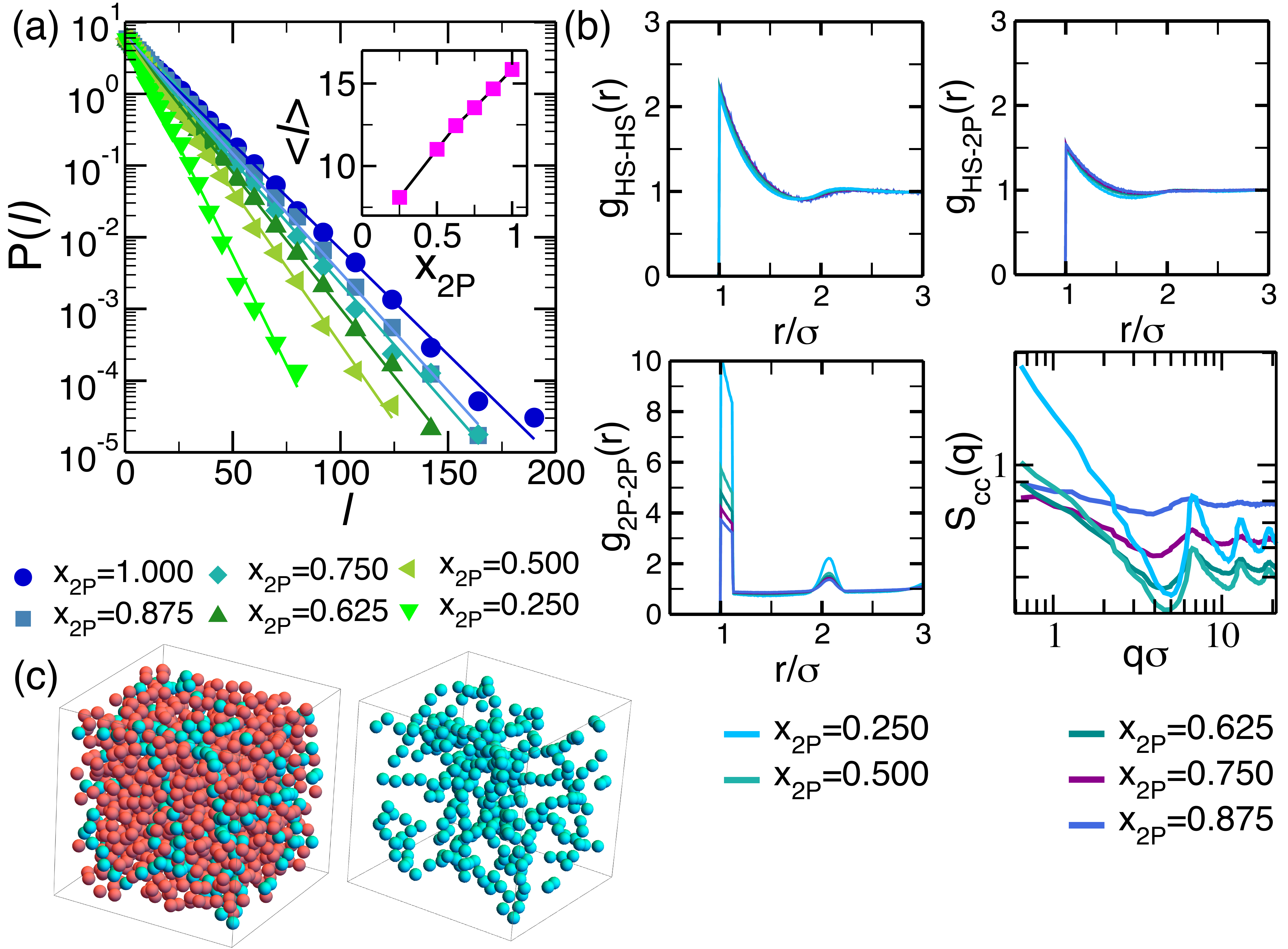}
    \caption{(a) Chain length $l$ distribution in mixtures of HS and 2P particles at $T=0.1$ and $\phi=0.262$. Inset: average chain length distribution $\langle l \rangle$ as a function of the 2P co-solute concentration $x_{\rm 2P}$ (b) Partial radial distribution functions $g_{\rm HS-HS}$(r) (upper-left panel), $g_{\rm HS-2P}$(r) (upper-right panel), and $g_{\rm 2P-2P}$(r) (lower-left panel) at $T=0.1$ and $\phi=0.262$. Lower right panel: concentration-concentration structure factor.
    (c) Left panel: snapshot of the mixture when $x_{\rm 2P}=0.25$ at $T=0.1$ and $\phi=0.262$. Red particles are HS, cyan particles are 2P. Right panel: same configuration, but only 2P particles are shown highlighting the formation of sparse chains. }
\label{fig:pl}
\end{figure*}

Before analyzing the effective interactions mediated by the binary mixture of co-solutes, it is worth to investigate the co-solute behaviour in the absence of large colloids. By fixing $\phi$ and $T$ and mixing 2P and HS together, we obtain an extra control parameter, the 2P concentration $x_{\textrm{2P}}$, which determines the 2P chain-length distribution $P(l)$ and its average value $\langle l\rangle$ as shown in Fig.~\ref{fig:pl}(a). We notice that, also in the case of binary mixtures, the chain length distribution is characterized by an exponential behaviour as predicted by the Flory-Stockmayer theory,~\cite{Flory1953} with an average length of the chain that increases on increasing $x_{\textrm{2P}}$. To gain insights into the structure of the mixture we calculate the partial radial distribution function with respect to all so-solute species, i.e., $g_{\rm 2P-2P}(r)$ and $g_{\rm HS-HS}(r)$, and that of the mixture, $g_{\rm HS-2P}(r)$ for different concentrations. 
The corresponding results are reported in Fig.~\ref{fig:pl}(b) revealing a striking difference between the different partial radial distribution functions.
In particular we find that the peak of $g_{\rm HS-HS}(r)$ is almost independent of concentration.
The same is observed for $g_{\rm HS-2P}(r)$.
On the contrary, we find that the first peak of $g_{\rm 2P-2P}(r)$ increases on increasing $x_{\rm 2P}$. To understand how the composition of the mixture affects the structure of the system we extract the average number of neighbours using the relation
\begin{equation}
N=4\pi\rho_i\int_0^{r_\textrm{min}}g_{ij}(r) r^2 \text{d}r,
\end{equation}
where the subscripts $i$ and $j$ indicates the species involved (2P-2P, HS-HS, HS-2P) and $r_{\rm min}$ represents the distance at which the first minimum after the first peak in $g_{ij}(r)$ is found. We find that the number of HS neighbours for one HS particle decreases with $x_{\textrm{2P}}=1-x_{\textrm{HS}}$
since it ranges from $\approx 9$ at $x_{\textrm{HS}}=1$ (only HS particles) to $\approx 1$ for $x_{\textrm{HS}}=0.125$ at the studied $\phi$. Analogously,  the number of 2P neighbours for a 2P particle changes from $\approx 6$ when $x_{\textrm{2P}}=1$ to $\approx 2$ at $x_{\textrm{2P}}=0.250$.
Interestingly, if we calculate the number of 2P neighbours of a 2P particle by choosing $r_{\textrm{min}}=\sigma+\delta$ we find that the number of neighbours does not change much as a function of $x_{\textrm{2P}}$.
The value is always close to $2$, which is a further indication of the fact that 2P particles are found mostly bonded with other 2P particles thus forming chains.
Finally, also from $g_{\rm HS-2P}(r)$ we find that the number of 2P neighbors of one HS particle decreases with $x_{\rm 2P}$ as expected. Due to the strong attraction between 2P particles, we are then interested to understand whether the two species could undergo demixing for some concentrations of 2P and HS particles.

\begin{figure*}[tbh]
\centering
  \includegraphics[width=0.75\linewidth]{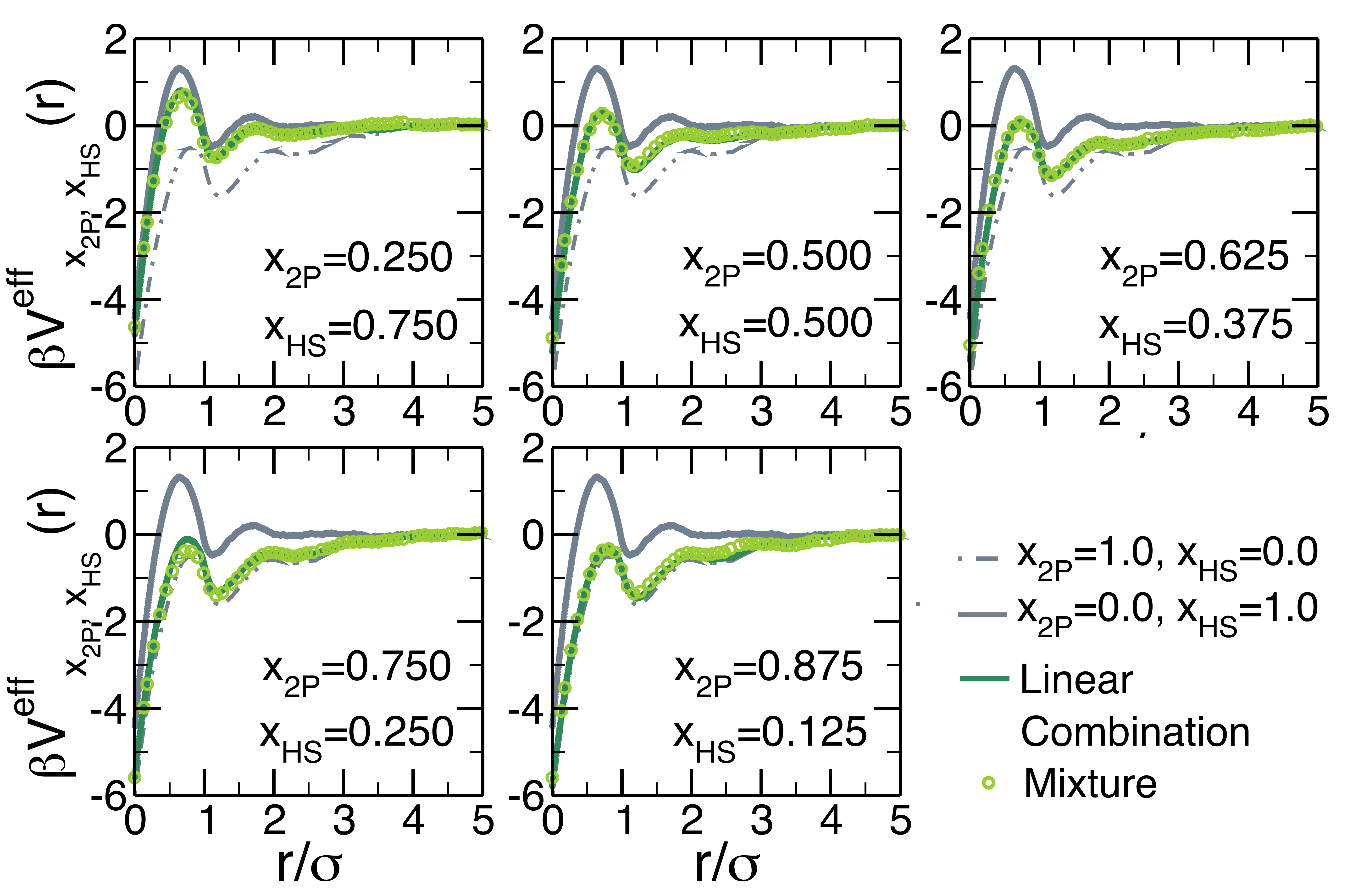}
    \caption{Effective potentials for different co-solute compositions at $T=0.1$ and $\phi=0.262$. Light green symbols show the effective potential mediated by the mixture.
    The grey curves show the effective potentials between two colloids immersed in a single-component co-solute of HS (upper curve) and 2P (lower curve) particles. The dark green curve is the linear combination of the two grey curves according to Eq.~\ref{eq:lincomb}.}
\label{fig:Veff2P-HS}
\end{figure*}

An indication of demixing is provided by the concentration-concentration structure factor $S_{cc}(q)$. In fact, close to the spinodal line there is usually a strong increase in concentration fluctuations within the mixture which is accompanied by a divergence of $\lim_{q\rightarrow 0}$ $S_{cc}(q)$.~\cite{Thornton1970}
The concentration-concentration structure factor for the 2P-HS mixture at different concentrations is 
reported in Fig.~\ref{fig:pl}(b) (lower-right panel). We observe that on decreasing $x_{\rm 2P}$, and hence on increasing $x_{\rm HS}$, weak oscillations appear in $S_{cc}(q)$, which progressively increase.
At the same time, the low $q$ part of the curve increases above $1$ for $x_{\rm 2P}=0.25$, signalling the appearance of inhomogeneities.
A closer look at the snapshots in Fig.~\ref{fig:pl}(c) helps to clarify what is happening in the binary mixtures.
As discussed above the radial distribution functions indicate that the 2P particles are always surrounded by at least two other 2P particles thus satisfying their bonds. At high $x_{\textrm{2P}}$ this can be achieved by allowing particles to be homogeneously distributed throughout the simulation box due to the high $\phi$ at which the system is simulated. On decreasing $x_{\textrm{2P}}$,  this is not possible anymore and to maximize the bonds 2P particles form few chains embedded in the sea of HS particles as shown in the snapshots.
The inhomogeneous distribution in the simulation box of 2P particles is indicated by the increase at small $q$ values of $S_{cc}(q)$.
Although this could be interpreted as demixing, we instead relate this result to the intrinsic behaviour that 2P particles have at low $T$ and low $\phi$. Indeed it has been shown that, under these conditions, the system of 2P particles can be treated as an ideal gas of chains, rather than particles, and an approximate expression of $S(q)$ at low $q$ values has been formulated starting from the $S(q)$ of a single chain.~\cite{SciortinoJCP2007}
Since the 2P chains are uniformly distributed throughout the box, we consider $x_{\rm 2P}=0.25$, $T=0.1$ and $\phi=0.262$ as a stable state point of the mixture.

\begin{figure*}
\centering
\includegraphics[width=0.65\textwidth]{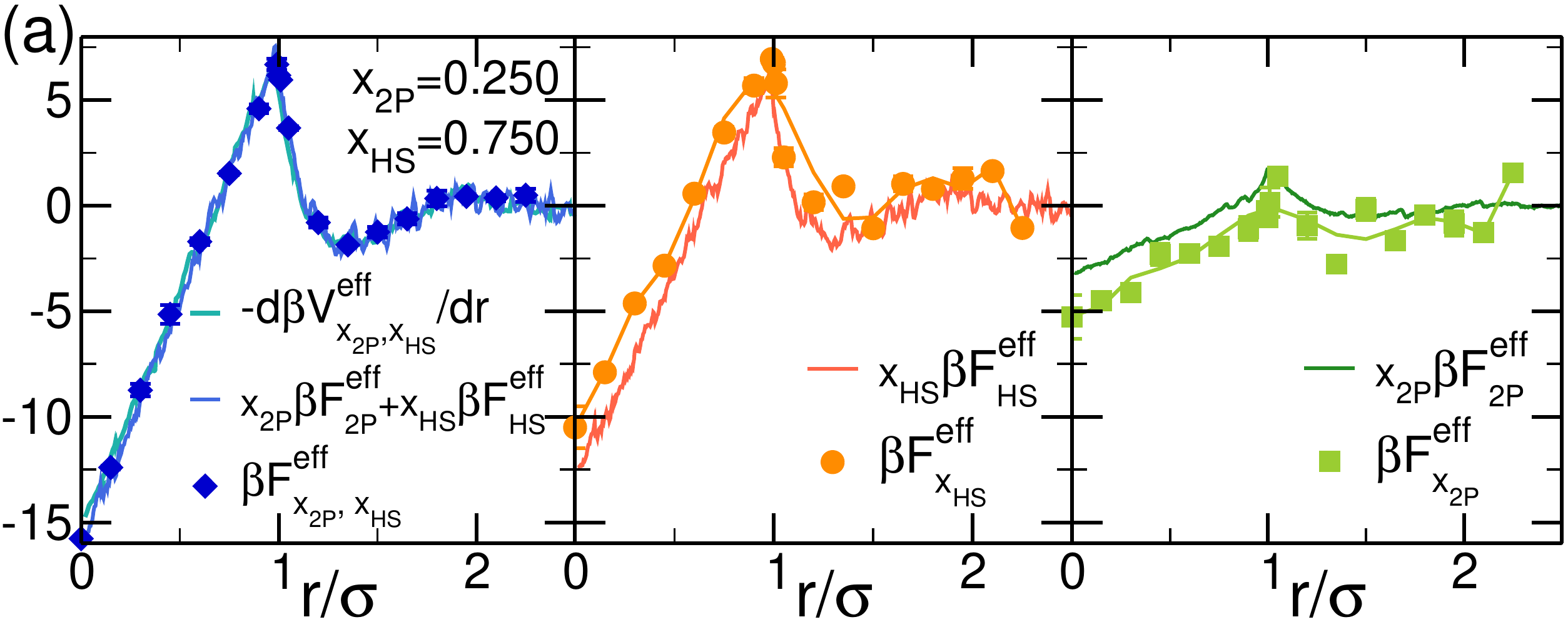}
\includegraphics[width=0.65\textwidth]{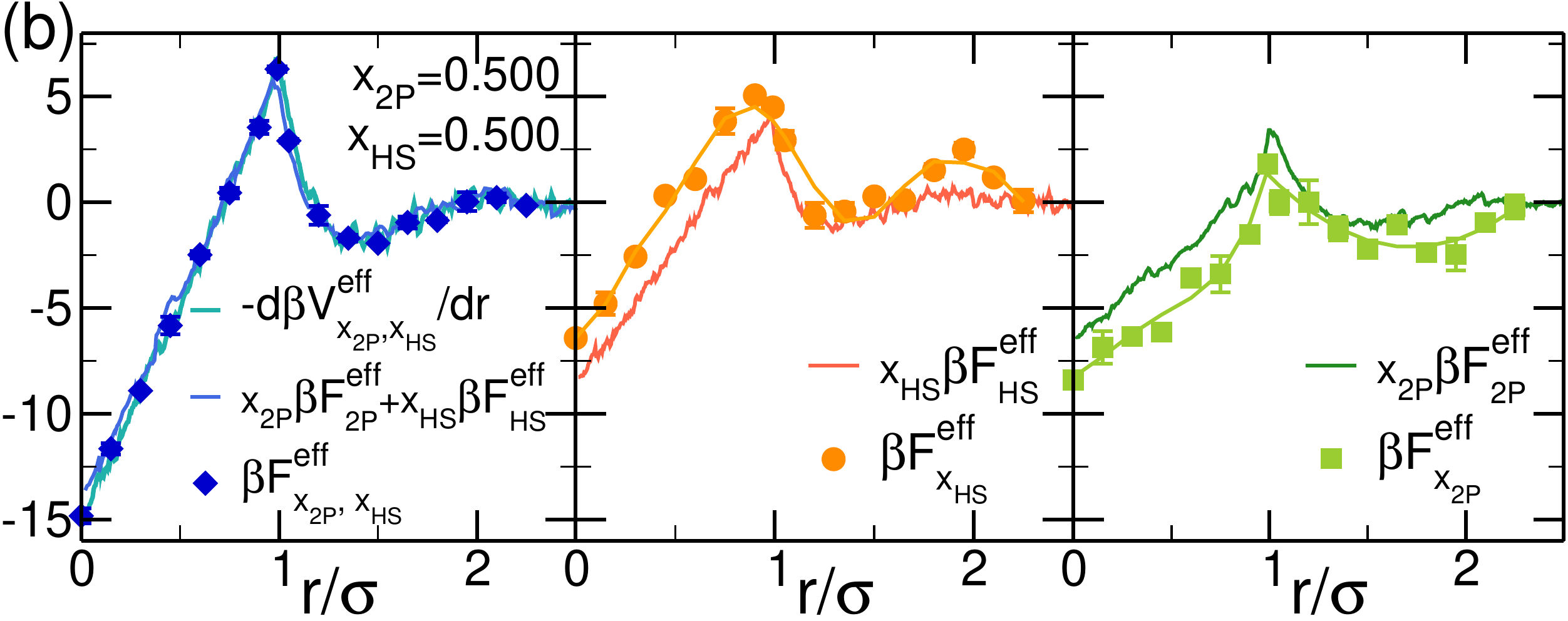}
\includegraphics[width=0.65\textwidth]{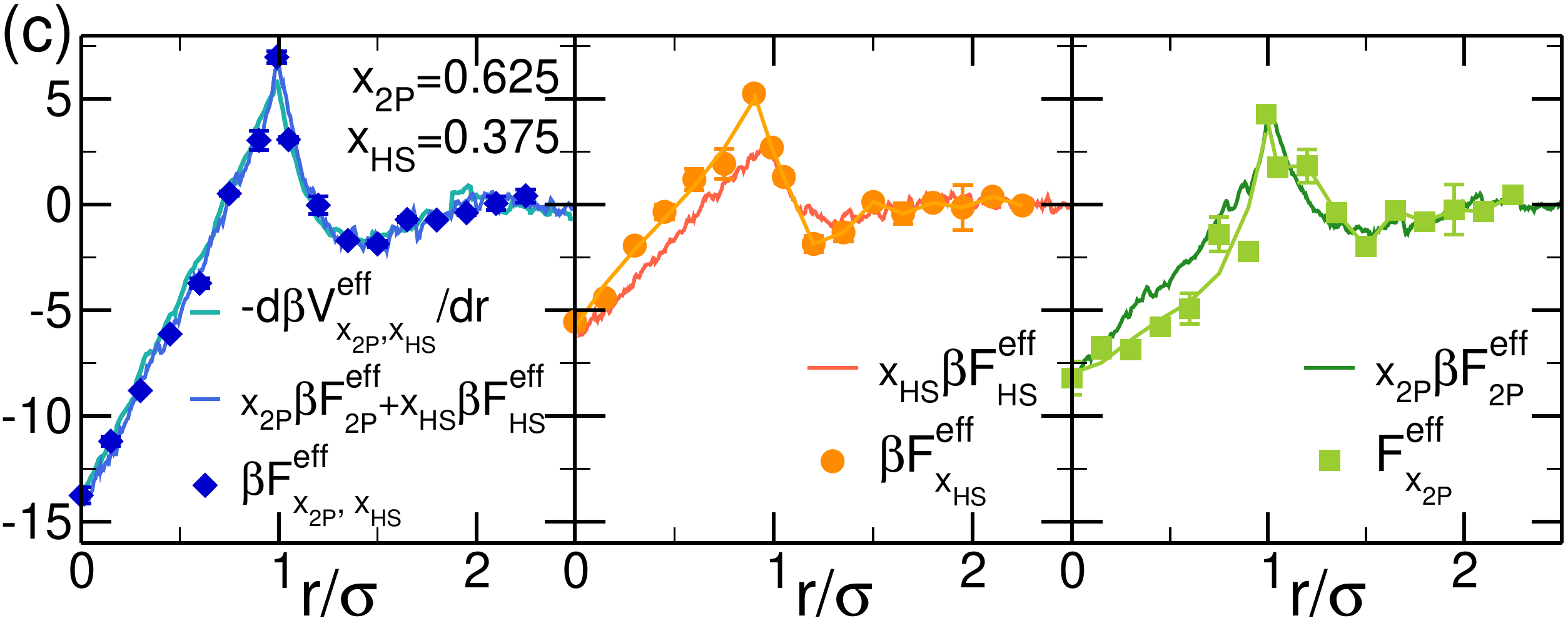}
\includegraphics[width=0.65\textwidth]{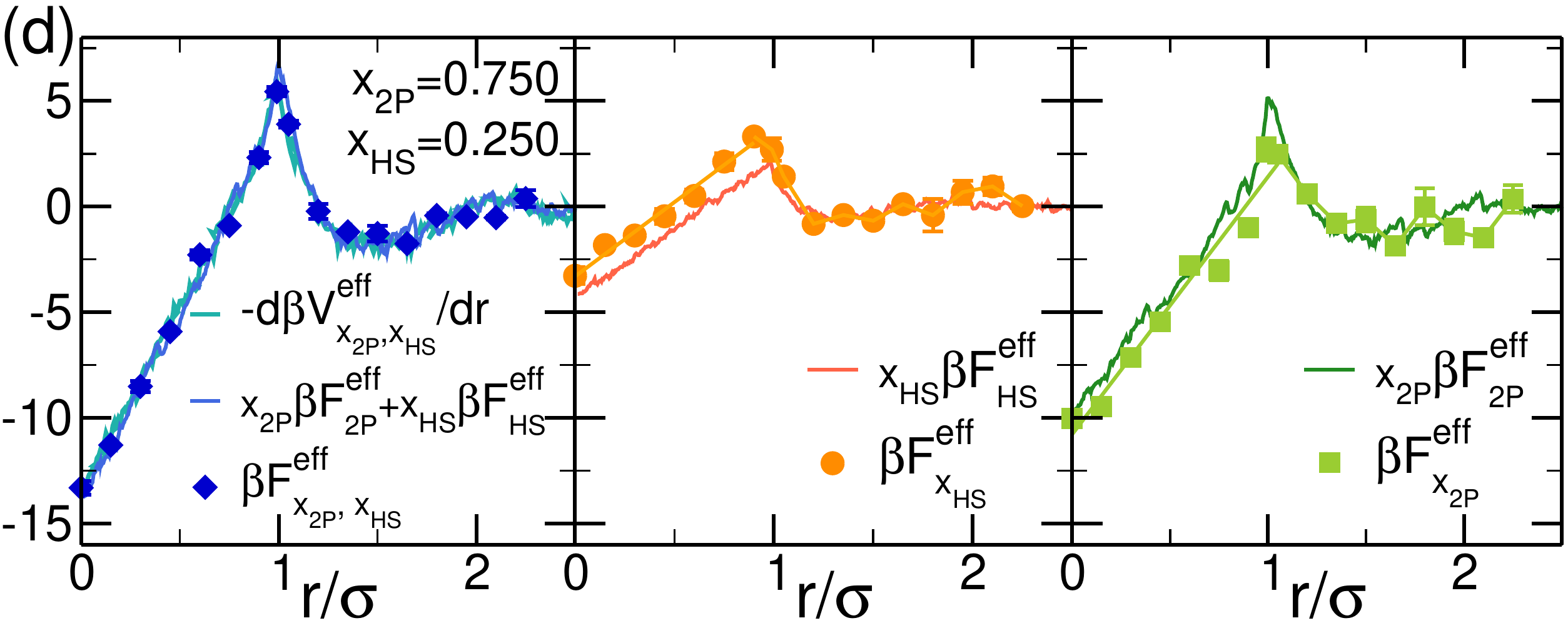}
\includegraphics[width=0.65\textwidth]{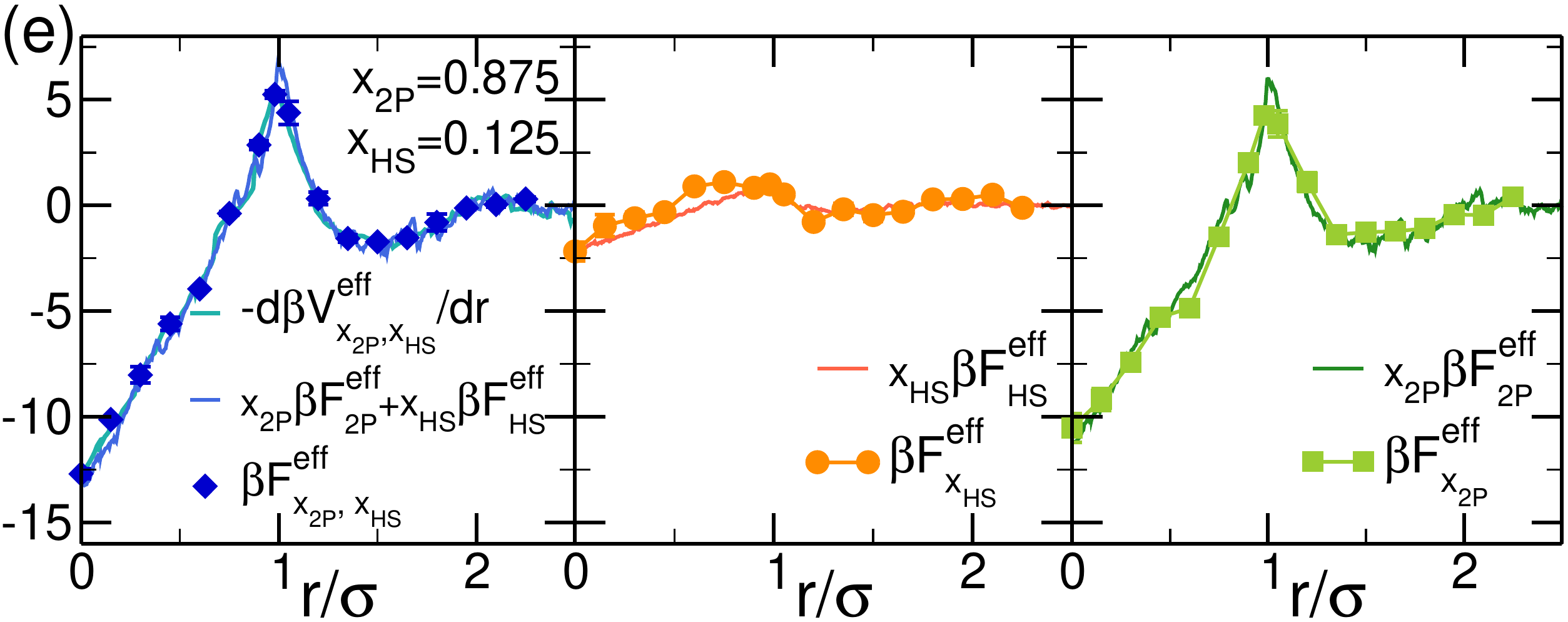}
\caption{Total effective force (left panels) and contribution to the force from the HS (central panels) and 2P (right panels) species of co-solute for a mixture of (a) $75\%$ of HS, $25\%$ of 2P (b) $50\%$ of HS, $50\%$ of 2P (c) $37.5\%$ of HS, $62.5\%$ of 2P (d) $25\%$ of HS, $75\%$ of 2P (e) $12.5\%$ of HS, $87.5\%$ of 2P. Symbols in central and right panels are interpolated with a polynomial function to provide a guide to the eye. The error bars represent the difference between the forces estimated from the overlaps of the left and right colloid with the co-solute.}
\label{fig:VirtualMoves2P}
\end{figure*}

\subsection{Effective potentials induced by HS-2P co-solute mixtures}
\label{sec:lincomb}
We focus our investigation on the state point with $T=0.1$ and $\phi=0.262$, since, as discussed above, for low enough $T$ and high enough $\phi$, the 2P particles have a tendency to form aligned chains. Under these conditions, the competition between bonding and excluded volume entropy leads to non-trivial effects in terms of effective interactions. Our aim is to understand whether or not the contribution to the effective potential stemming from the two species can be simply obtained from the linear combination of the effective potentials  of the single component co-solutes. In other words, given the effective potential between two large HS colloids immersed in a mixture of 2P and HS co-solute with concentration $x_{\rm 2P}$ and $x_{\rm HS}=1-x_{\rm 2P}$, i.e., $V^{\textrm{eff}}_{x_{\rm 2P},x_{\rm HS}}(r)$, we want to verify whether
\begin{equation}\label{eq:lincomb}
\beta V^{\textrm{eff}}_{x_{\rm 2P},x_{\rm HS}}(r)= x_{\rm 2P} \beta V^{\textrm{eff}}_{\rm 2P}(r) + x_{\rm HS}\beta V^{\textrm{eff}}_{\rm HS}(r)
\end{equation}
holds. We therefore compare the effective potential of the two colloids immersed in the mixture, for different concentrations of the two species, with the corresponding ones calculated from Eq.~\ref{eq:lincomb}. Results are shown in Fig.~\ref{fig:Veff2P-HS}.
The grey curves refer to \veff{} calculated for the single component HS and 2P co-solutes and are used as reference potentials. Notice, that on increasing the concentration of 2P particles, the effective potential becomes totally attractive as expected. In addition, for all studied concentrations, an almost perfect agreement is found with the potential obtained from Eq.~\ref{eq:lincomb}, suggesting the additivity of the total effective interactions of the mixture.

On the basis of this result for the total interactions, we would also expect that each contribution to the total potential, arising from the two species in the mixture separately, would coincide with the corresponding single-component ones. To shed light on this point, we cannot rely on
the umbrella sampling method, which does not provide the two terms separately, but we resort to 
the \emph{virtual displacements} method explained in Sec.~\ref{sec:modelsmethods}.
This method allows us to have an independent measurement of the effective force between the two colloids, which we compare with that obtained from the umbrella sampling method via the relation
\begin{equation}
    \beta F^{\rm eff}_{ x_{\rm 2P}, x_{\rm HS}}(r)=
    -\frac{\text{d}}{\text{d}r}\beta V^{\rm eff}_{x_{\rm 2P}, x_{\rm HS}}(r).
\end{equation}
In addition, and most importantly, the \emph{virtual displacements} method provides the effective forces stemming from the two components of the mixture separately. Fig.~\ref{fig:VirtualMoves2P} (left panels) shows the total effective force mediated by the mixtures compared with the force calculated from the umbrella sampling method and that obtained from the linear combination of Eq.~\ref{eq:lincomb} (where, instead of $V_{\textrm{eff}}$ we use $F_{\textrm{eff}}$). Again, we find confirmation that the agreement is very good, independently of the composition of the mixture.

In addition, the central and left panels of Fig.~\ref{fig:VirtualMoves2P} report the components of the effective force due to the HS species ($\beta F^{\textrm{eff}}_{x_{\textrm{HS}}}$) and to the 2P species
($\beta F^{\textrm{eff}}_{x_{\textrm{2P}}}$), respectively. Both forces are compared with the effective forces stemming from the single component co-solutes, multiplied by the concentration of the species, i.e., $x_{\textrm{HS}}\beta F^{\textrm{eff}}_{\textrm{HS}}$ and $x_{\textrm{2P}}\beta F^{\textrm{eff}}_{\textrm{2P}}$, respectively, which are indeed the two terms in Eq.~\ref{eq:lincomb}.
Interestingly, we find that the separate contributions of the two species do not match their corresponding term in Eq.~\ref{eq:lincomb}, even if their sum satisfies the same equation. In particular, while the HS force $\beta F^{\textrm{eff}}_{x_{\textrm{HS}}}$  is larger than $[x_{\textrm{HS}}\beta F^{\textrm{eff}}_{\textrm{HS}}]$, the 2P force $\beta F^{\textrm{eff}}_{x_{\textrm{2P}}}$ is smaller than $[x_{\textrm{2P}}\beta F^{\textrm{eff}}_{\textrm{2P}}]$.
This result is found for all mixture compositions,  but the deviations are largest at $x_{\textrm{2P}}=0.5$ and significantly decrease with increasing 2P concentration.
For $x_{\textrm{2P}}=0.875$ they become negligible.

\begin{figure*}
\centering
\includegraphics[width=1.0\textwidth]{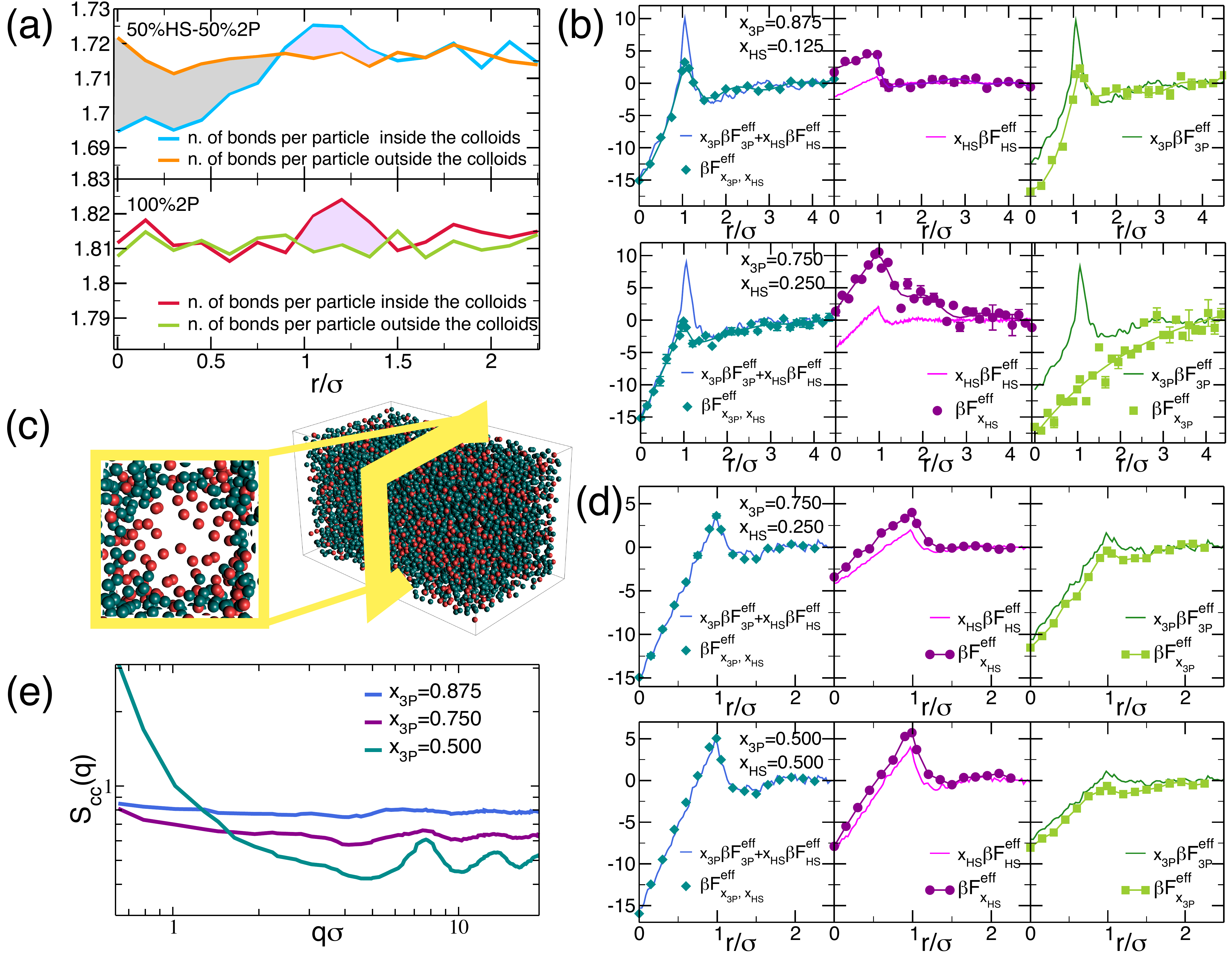}
\caption{(a) Upper Panel: average number of bonds for 2P particles at contact with the colloids in a co-solute mixture with $50\%$ of HS and $50\%$ of 2P particles at $T=0.1$ and $\phi=0.262$. Lower Panel: the same but for a single-component co-solute at the same $T$ and $\phi$.
(b) Total effective force (left panels) and contribution to the force from the HS (central panels) and 3P (right panels) species of co-solute for a mixture of  (upper panels) $12.5\%$ of HS and $87.5\%$ of 3P (lower panels) $25\%$ of HS and $75\%$ of 3P at $T=0.125$ and $\phi=0.262$. Symbols in central and right panels are interpolated with polynomials to provide a guide to the eye. (c) Snapshot of the mixture made of $25\%$ HS  and $75\%$ 3P particles. Two colloids are placed in the middle of the simulation box at a distance $r/\sigma=1.2$. A magnification of the snapshot in between the two colloids shows that only HS particles (in red) are found there, while 3P particles (in green) are excluded outside.
(d) Total effective force (left panels) and contribution to the force from the HS (central panels) and 3P (right panels) species of co-solute for a mixture of  (upper panels) $25\%$ of HS and $75\%$ of 3P (lower panels) $50\%$ of HS and $50\%$ of 3P at $T=0.2$ and $\phi=0.262$. Symbols in central and right panels are interpolated with polynomials to provide a guide to the eye.
(e) Concentration-concentration structure factors for a mixture of HS and 3P particles at $T=0.125$ and $\phi=0.262$ for different concentrations of the two species.
}
\label{fig:AvBonds}
\end{figure*}

Since the \emph{virtual displacements} method is based on counting the number of overlaps of the two colloids with the co-solute in between and outside them, the behaviour of the two components of the effective force indicates that there are more HS co-solute particles between the two colloids than expected from $x_{\textrm{HS}}\beta F^{\textrm{eff}}_{\textrm{HS}}$. Consequently, there are less 2P co-solute particles in this region than expected from $x_{\textrm{2P}}\beta F^{\textrm{eff}}_{\textrm{2P}}$. This is true for almost all concentrations when the colloids are at a distance $r/\sigma\lesssim 1$. For $x_{\textrm{2P}}\geq 0.5$ this also occurs at larger distances.
While the HS co-solute particles satisfy only entropic constraints, the 2P co-solute  behaviour is always determined by the competition of bonding energy and excluded volume.
In particular, we expect that the lower number of 2P particles  is related to a reduced ability to form bonds in between the two colloids (at least when $r/\sigma\lesssim 1$). Thus, an increasing concentration of HS particles in this region makes the 2P particles stay outside the colloids where they still are able to form bonds. 
To test this hypothesis, we calculate the average number of bonds of 2P particles that are in contact with the colloids, i.e., considering the 2P particles exerting a force on the HS colloids.
The results for this observable are shown in Fig.~\ref{fig:AvBonds}(a) as a function of the colloid's surface-to-surface distance $r$ distinguishing the 2P particles in two groups: those  on the colloids from the outside and those exerting a force from the in-between region, as determined during the \emph{virtual displacements} method (cf. Sec.~\ref{sec:modelsmethods}).
The calculation has been performed for the co-solute mixture with $50\% \textrm{HS} -50\% \textrm{2P}$ (upper panel), where this effect should be strongest.
We compare the results to those obtained for the single-component 2P co-solute at the same packing fraction and temperature (lower panel).
The grey area drawn between the two curves in the upper panel highlights that there is a gap between the number of bonds per particle in the two regions.
Although the difference is small, the gap is clearly absent in the case of the single-component 2P co-solute, indicating that HS particles influence the ability of 2P particles to form bonds in the region between the two colloids. Finally, for both cases we also highlight (purple area) a small region for $r/\sigma>1$ where the number of bonds formed between the two colloids is larger than outside them. This is related to the nematization effect discussed in Sec.~\ref{sec:mono}: when the depletion zone disappears and 2P particles can access to the region between the colloids they can satisfy both entropic and energetic contraints by reorienting their bonds in order to form parallel chains.
 
\subsection{Effective potentials induced by HS-3P co-solute mixtures}
\label{sec:3P-HS}

We repeat the analysis discussed above for the case of two HS colloids immersed in a mixture of co-solute particles made of HS and 3-patch (3P) particles. Previous work~\cite{garcia2017effective} has shown that the effective interactions generated by 3P co-solute particles at high $\phi$ and low $T$ exhibit the same features as the 2P ones, i.e., 3P particles give rise to an oscillatory effective attraction between HS colloids.
Again, the attractive oscillations are related to the orientation of particle bonds, which, in this case, tend to form planar sheets of 3P particles under confinement which are orthogonal to the vector that joins the centres of the two colloids. Differently from 2P particles, the phase diagram of 3P particles contains a gas-liquid coexistence region with a critical point that, for the model parameters employed in this work, has been located at $\phi_c=0.072$ and $T_c=0.125$.~\cite{GnanSM2016} 
Here we investigate the mixture of HS and 3P particles at $T=0.125$ and $\phi=0.262$ for which the effective interaction mediated by the two species separately is already known.~\cite{garcia2017effective}
Fig.~\ref{fig:AvBonds}(b) shows the total effective force and the contribution to the force from 3P and HS co-solutes at two different concentrations of the two species ($x_{\textrm{3P}}=0.875$ and $x_{\textrm{3P}}=0.750$). Interestingly, in this case, upon increasing the concentration of HS, we find a progressive deviation from the additive behavior of the total force. In addition, the two components of the effective force display a long-range tail which clearly signals the presence of some kind of correlation between the co-solutes. It is worth to stress that, at the two studied compositions, the mixtures are stable with respect to demixing in the absence of the two colloids, but are not too far from the onset of this transition. 
This is shown in Fig.~\ref{fig:AvBonds}(e), where demixing clearly takes place for $x_{\textrm{3P}}=0.500$, as signalled by the low-$q$ divergence of the concentration-concentration structure factor.  Thus, the results obtained for the effective forces can be interpreted in terms of a local phase separation driven by the presence of the two colloids in the mixture. This hypothesis, also explaining the long-range components of the effective force, is supported by the snapshot shown in Fig.~\ref{fig:AvBonds}(c) of the mixture in the presence of the HS colloids, where the region between the two colloids is devoid of 3P particles even for $r/\sigma>1$. Consequently, only the HS co-solute particles are found in this region, indicating local demixing. 

Finally, to show that such non-additive behaviour disappears far from demixing, we have studied a 3P-HS cosolute mixture at a higher temperature ($T=0.2$) at the same packing fraction. The results are shown in Fig.~\ref{fig:AvBonds}(d) for two different mixture compositions. As for the 2P and HS mixtures, in this case the total effective force coincides with that arising from the linear combination of the two single-component contributions. In addition, the effective force generated by HS co-solute particles is always higher than the expected one and it is balanced by that of the 3P particles, which is smaller than expected, in agreement with the findings for the 2P and HS mixture results, shown in Fig.~\ref{fig:VirtualMoves2P}. Also in the present case, the two effects cancel out such that the total force is found to be additive.

\section{Conclusions}
\label{sec:conclusions}

In this work we have investigated the effective interactions of two HS colloids immersed in a binary mixture of smaller co-solute particles.
The binary mixtures considered contain HS and patchy particles (2P or 3P). 2P particles are known to reversibly assemble into chains and a previous study on the effective interactions between colloids mediated by those particles reported that the competition between bonding  and entropic constraints modulates the effective potential giving rise to  attractive oscillations between the colloids.~\cite{garcia2017effective} This occurs mainly at low $T$ and high $\phi$ conditions, where the present analysis is mostly focused.
When a mixture of 2P and HS co-solutes is considered, the effective potential mediated by the co-solute shows a continuous transition from a standard effective potential found in highly asymmetric HS binary mixtures, to a completely attractive potential as observed for HS colloids in the pure 2P co-solute. Despite the non-trivial effects driven by the assembly of 2P particles and their confinement close to the colloids, we find that, when two HS colloids in a mixture of HS and 2P co-solutes are considered, the effective interactions of the mixture are always additive and, thus, they can simply be estimated from the linear combination of the effective interactions mediated by the pure HS and the pure 2P co-solute.
Interestingly, we find that the separate contributions to the effective force from both HS and 2P particles deviate from the expected ones, but these deviations cancel out providing a total effective force that satisfies the linear combination.
This is attributed to the fact that a larger number of 2P particles are found outside the two colloids than in the confined region between them, and viceversa for the HS particles. We have shown that this is related to the fact that 2P particles are able to form more bonds outside the colloids while HS particles are better accommodated in the confined region between the two colloids when their surface-to-surface distance is $r<\sigma$.

We have repeated the same analysis for HS-3P co-solute mixtures finding that, for some concentrations the linear combination does not work to a satisfactory degree. This behaviour can be attributed to local demixing induced by the presence of the colloids, despite no signature of demixing occurs in the mixture without the HS colloids, even if the considered state point is close to the demixing line. When the two colloids are placed in the mixture, we find that 3P particles are fully depleted from the confined region between the colloids even for $r>\sigma$. However, if the  temperature is increased, linear combination yields again the correct result also for the HS-3P mixtures.

In conclusion, we have presented a first survey of effective interactions mediated by binary co-solute mixtures in which one component is capable of self-assembly. The present findings open the way to use additive effective interactions in the presence of binary co-solutes. Indeed, if the linear combination works, the knowledge of the potential in the single-component co-solutes will be sufficient to predict the total effective potentials in the mixtures. 
This approach provides a promising route for the inverse design of new kinds of colloid-colloid interactions \emph{in silico}, able to induce peculiar self-assembly or collective behavior in the colloids themselves, using a method that is, in principle, readily realizable also in experiments. The study of different types of co-solutes (e.g., inverse patchy colloids~\cite{bianchi2017inverse}) as well as of state-of-the-art DFT~\cite{RothPRE62,gnan2012DFT,davidchack2016hard,teixeira2019patchy,stopper2020remnants} and recent integral equation theory~\cite{Cummings2020,Cummings2021} calculations will surely provide additional insights in this direction.

\begin{acknowledgments}
P.H.H. thanks the Austrian Science Fund FWF for support (Erwin Schrödinger Fellowship No. J3811 N34)
\end{acknowledgments}

\subsection*{AVAILABILITY OF DATA}
The data that support the findings of this study are available from the corresponding author upon reasonable request.

\bibliography{bib.bib}

\end{document}